\documentclass[10pt,twocolumn,letterpaper]{article}
\usepackage{spconf,amsmath,graphicx}
\usepackage{times}
\usepackage[caption=false]{subfig}
\usepackage{amssymb}
\usepackage{algorithm,algorithmic}
\usepackage{multirow}
\usepackage[american]{babel}
\usepackage[pagebackref=true,breaklinks=true,colorlinks,bookmarks=false]{hyperref}
\usepackage{array}
\usepackage{enumitem}

\newcolumntype{P}[1]{>{\centering\arraybackslash}p{#1}}
\newcolumntype{M}[1]{>{\centering\arraybackslash}m{#1}}
\title{Enhancing HEVC compressed videos with a partition-masked convolutional neural network}
\name{Xiaoyi He$^{1}$ \quad Qiang Hu$^{1}$ \quad Xintong Han$^{2}$ \quad Xiaoyun Zhang$^{1}$ \quad Chongyang Zhang$^{1}$ \quad Weiyao Lin$^{1^{\star}}$\thanks{Copyright 2018 IEEE. Published in the IEEE 2018 International Conference on Image Processing (ICIP 2018), scheduled for 7-10 October 2018 in Athens, Greece. Personal use of this material is permitted. However, permission to reprint/republish this material for advertising or promotional purposes or for creating new collective works for resale or redistribution to servers or lists, or to reuse any copyrighted component of this work in other works, must be obtained from the IEEE. Contact: Manager, Copyrights and Permissions / IEEE Service Center / 445 Hoes Lane / P.O. Box 1331 / Piscataway, NJ 08855-1331, USA. Telephone: + Intl. 908-562-3966.}}

\address{$^{1}$ Department of Electronic Engineering, Shanghai Jiao Tong University, China \\
    $^{2}$Department of Electrical and Computer Engineering, University of Maryland \\
    ($^{\star}$Corresponding Author: wylin@sjtu.edu.cn)}

\begin{document}
%
\maketitle
\begin{abstract}
In this paper, we propose a partition-masked Convolution Neural Network (CNN) to achieve compressed-video enhancement for the state-of-the-art coding standard, High Efficiency Video Coding (HECV). More precisely, our method utilizes the partition information produced by the encoder to guide the quality enhancement process. In contrast to existing CNN-based approaches, which only take the decoded frame as the input to the CNN, the proposed approach considers the coding unit (CU) size information and combines it with the distorted decoded frame such that the degradation introduced by HEVC is reduced more efficiently. Experimental results show that our approach leads to over 9.76\% BD-rate saving on benchmark sequences, which achieves the state-of-the-art performance.

\end{abstract}
\section{Introduction and related work}
\label{sec:intro}

Recently, the fast development of video capture and display devices has brought a dramatic demand for high definition (HD) contents.
High Efficiency Video Coding (HEVC) \cite{hevc} provides higher compression performance compared to the previous standard H.264/AVC by 50\% of bitrate saving on average at a similar perceptual image quality \cite{Comparison_of_the_coding}. However, HEVC videos still contain compression artifacts, such as blocking artifacts, ringing effects, blurring, etc.. Therefore, it is desired to study on improving the visual quality of the decoded videos.


Recently, many deep learning based approaches \cite{arcnn,vrcnn,qecnn,ifcnn} have been proposed to enhance the visual quality of compressed images and videos. \cite{ifcnn} designed a CNN to replace the loop filter \cite{deblocking,sao} in HEVC. \cite{arcnn} developed an Artifacts Reduction CNN (ARCNN) built upon \cite{srcnn}, which reduces the JPEG compression artifacts. Following \cite{arcnn}, \cite{vrcnn} and \cite{qecnn} proposed a Variable-filter-size Residual-learning CNN (VRCNN) and a Quality Enhancement CNN (QECNN) respectively as post-processing techniques to further improve the quality of the compressed videos in HEVC. However, existing works only consider the appearance of input coding units (CUs) or frames, while the partition variations in different CUs and frames are neglected. In practice, since the partition information (e.g., 16$\times$16, 8$\times$8) is introduced by the block-wise processing and quantization of HEVC, this indicates the source of visual compression artifacts. Thus, we use the partition information to effectively guide the quality enhancement process performed by CNN.

 \begin{figure}[t]
   \centering
   \includegraphics[width=0.48\textwidth]{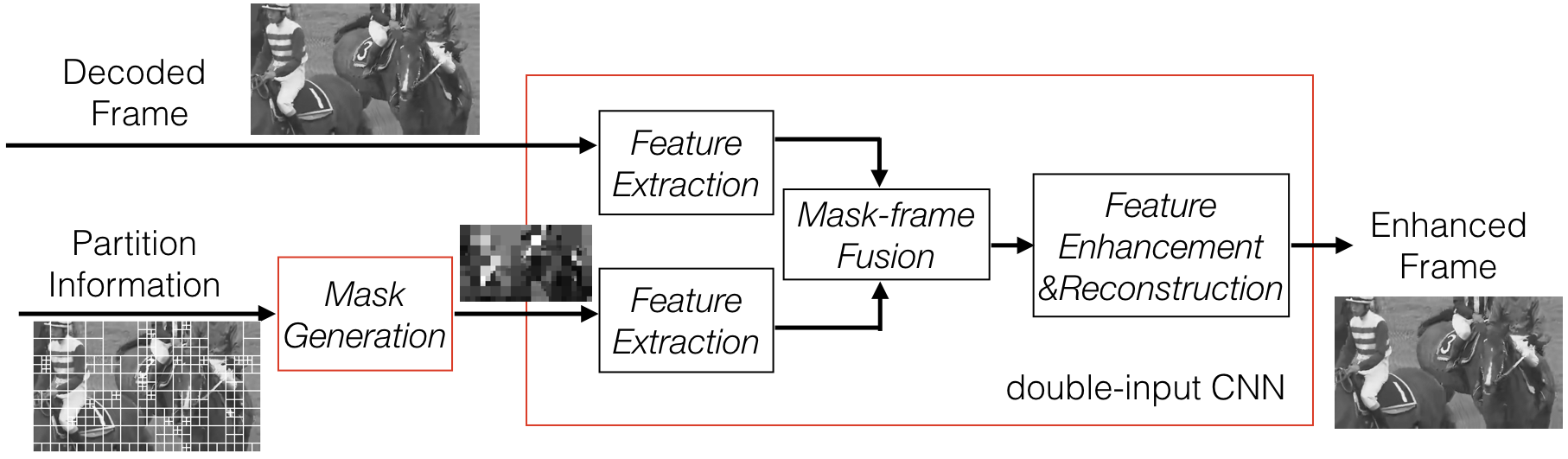}
   \caption{Overview of the proposed framework.}
   \label{fig:base_framework}
 \end{figure}

To this end, we propose a novel approach which first derives a carefully designed mask from a frame's partition information, and then uses it to guide the quality enhancement process of the decoded frame through a double-input CNN. As a result, the visual quality of HEVC-compressed videos can be more properly improved under the same bit rate. The diagram of the proposed approach is shown in Fig. \ref{fig:base_framework}. In summary, our contributions are 3 folds:

\begin{enumerate}[noitemsep]
\item We develop a novel framework that utilizes the partition information to guide the CNN-based quality enhancement process in HEVC, where a mask derived from a decoded frame's partition information is fused with this decoded frame through a double-input CNN to accomplish quality enhancement.
\item Under this framework, we systematically investigate different mask generation and mask-frame fusion methods and find the best strategies. We also demonstrate that our approach is general and can be integrated into the existing HEVC compressed-video enhancement methods to further improve their performances.
\item We establish a large-scale dataset which contains 202,251 training samples for training reliable compressed-video enhancement models. This dataset will be made publicly available to facilitate further research.
\end{enumerate}


\section{Overview of our approach}
\label{sec:overview}
The framework of our approach is shown in Fig. \ref{fig:base_framework}. Each decoded frame and its corresponding mask, which is generated using the frame's partition information (cf. mask generation in Fig. \ref{fig:base_framework}), are fed to a double-input CNN. Inside this CNN, the features of the mask and decoded frame are first extracted through two individual streams and then fused into one (cf. mask-frame fusion). The rest layers of the double-input CNN perform the feature enhancement, mapping, reconstruction, and output the quality-enhanced decoded frame.

\begin{figure}[t]
\centering
\subfloat[original image with partition information]{\includegraphics[width=0.32\linewidth]{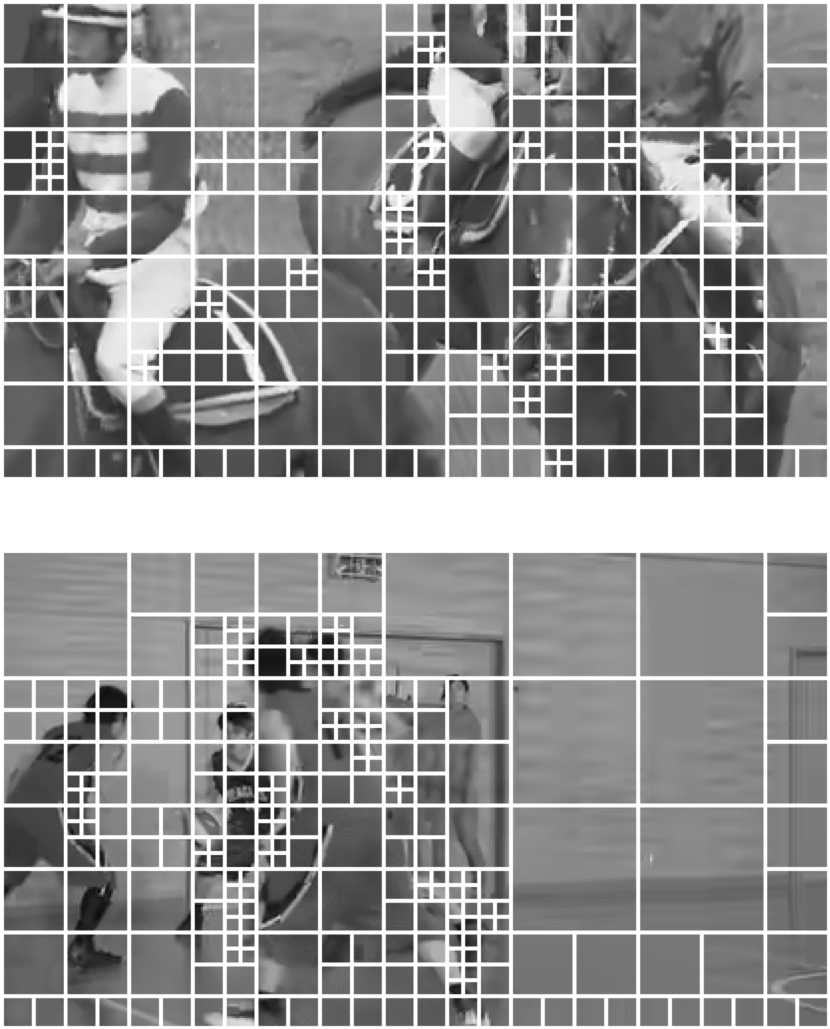}\label{fig:3a}}
\hfill
\subfloat[Mean-based mask]{\includegraphics[width=0.32\linewidth]{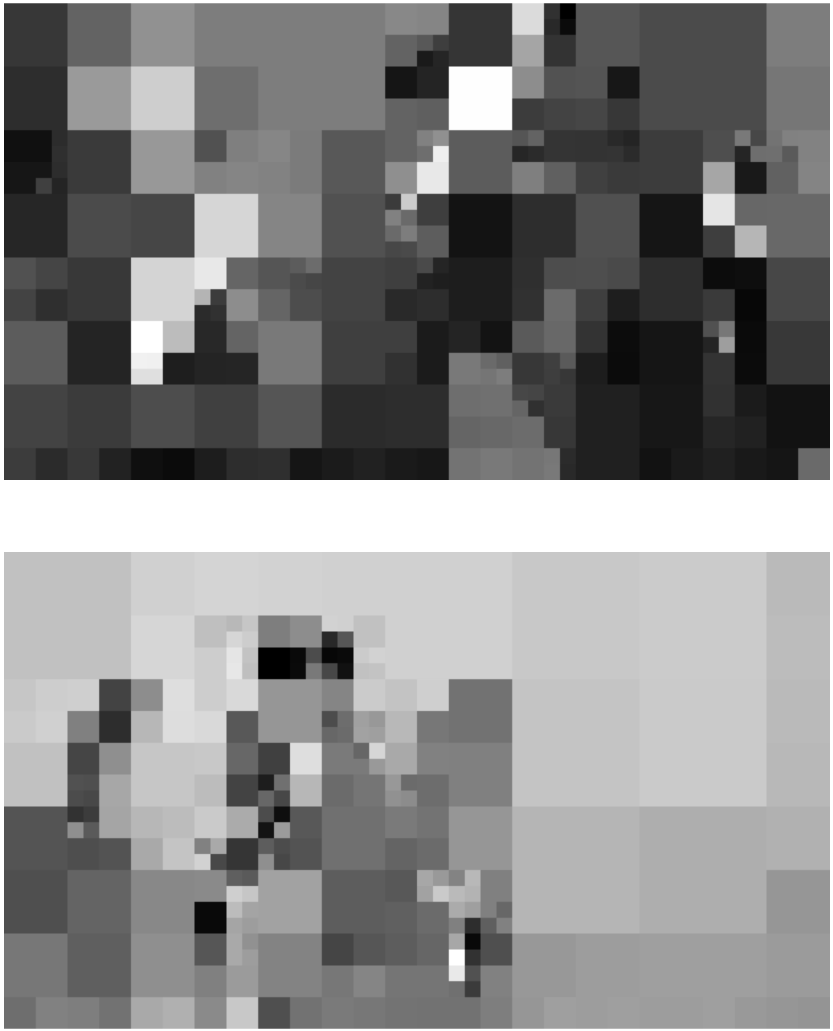}\label{fig:3b}}
\hfill
\subfloat[Boundary-based mask]{\includegraphics[width=0.32\linewidth]{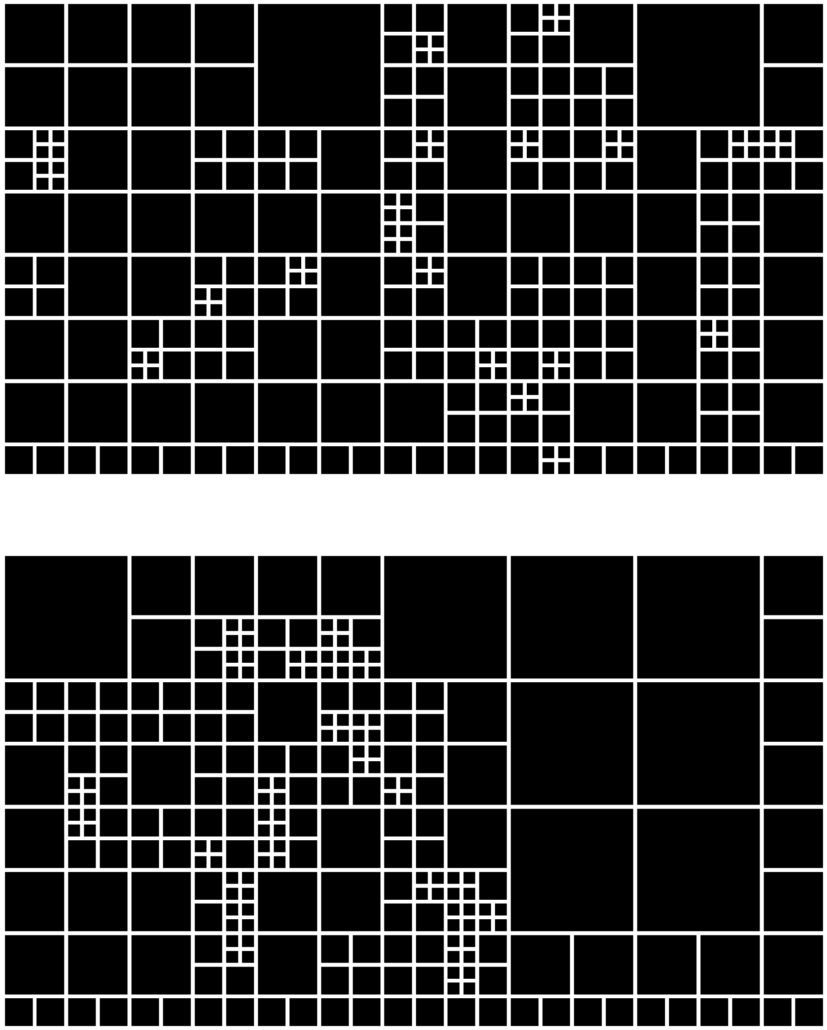}\label{fig:3c}}%
\hfill
\caption{Two Examples of Boundary-based mask and Mean-based mask.}
\label{fig:mask}
\end{figure}

\section{The proposed approach}
\label{sec:detail}
In this section, we first discuss the key components of our approach -- mask generation and mask-frame fusion strategies. Then, we describe the proposed double-input CNN.

\subsection{Mask generation and mask-frame fusion strategies}
  \label{sec:mask_and_fusion}
 Since the block-wise transform and quantization are performed in HEVC during encoding, the quality degradation of compressed frames is highly related to the coding unit splitting. Thus, the partition information contains useful clues for eliminating the artifacts present during the encoding. Considering this, we design a mask based on the partition information of CUs to guide the quality enhancement process.

\begin{figure}[t]
    \centering
  \subfloat[]{\includegraphics[width=0.48\linewidth]{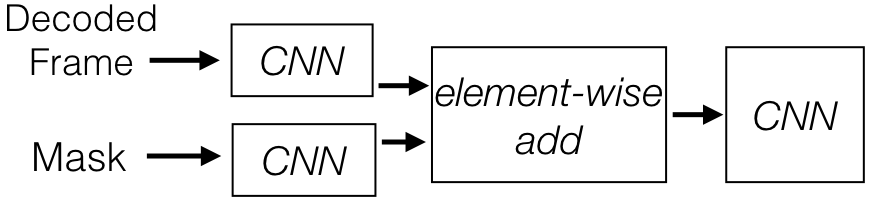}\label{fig:4a}}
  \hfill
  \subfloat[]{\includegraphics[width=0.48\linewidth]{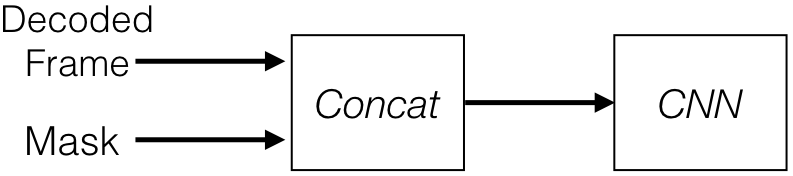}\label{fig:4b}}
  \hfill
  \subfloat[]{\includegraphics[width=0.5\linewidth]{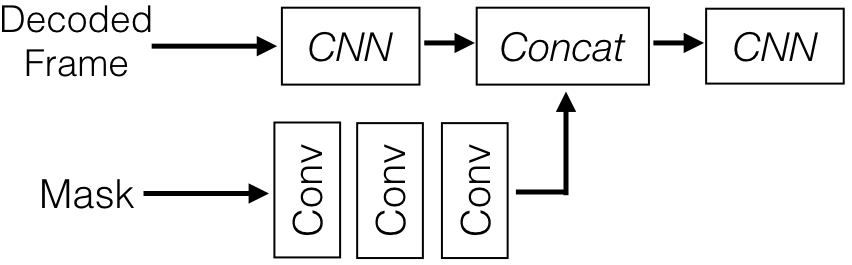}\label{fig:4c}}
  \hfill
  \caption{(a) Add-based fusion. (b) Concatenate-based fusion. (c) Early fusion.}
  \label{fig:fusin_str}
\end{figure}

\textbf{Generation of the mask}. We introduce two strategies to generate masks from an HEVC-encoded frame's partition information:

 \begin{itemize}[noitemsep]
  \item Mean-based mask (MM). We fill each partition block in a frame with the mean value of all decoded pixels inside this partition. An example of a generated mean-based mask is shown in Fig. \ref{fig:3b}. As we can see that the different partition blocks are properly displayed in the mask. In this way, when we fuse it with the decoded frame during the enhancement process, it can effectively distinguish different partition modes and reduce the compression artifacts more effectively.
  \item Boundary-based mask (BM). We also introduce a boundary-based mask generation strategy. In this boundary-based mask, the boundary pixels between partitions are filled with value 1 and the rest non-boundary pixels are filled with value 0, as shown in Fig. \ref{fig:3c}. The width of the boundary is set to 2.
\end{itemize}

 \textbf{Mask-frame fusion strategies}. As we mentioned in Section \ref{sec:overview}, the mask is fed to CNN and integrated with its corresponding decoded image to get the fused feature maps. We also introduce three strategies to fuse the information of a decoded frame and its corresponding mask:

\begin{itemize}[noitemsep]
  \item Add-based fusion (AF). As shown in Fig. \ref{fig:4a}, we first extract the feature maps of the mask using CNN and then combine it with the feature maps of the input frame using element-wise add layer.
  \item Concatenate-based fusion (CF). We concatenate the mask and frame as the input to the CNN. Then the two-channel image is fed to CNN directly as shown in Fig. \ref{fig:4b}.
  \item Early fusion (EF). We extract the features of mask only using three convolutional layers and integrate it into the network as shown in Fig. \ref{fig:4c}.
\end{itemize}

\begin{figure*}[t]
  \centering
  \subfloat[]{\includegraphics[width=0.7\textwidth,height=0.23\textwidth]{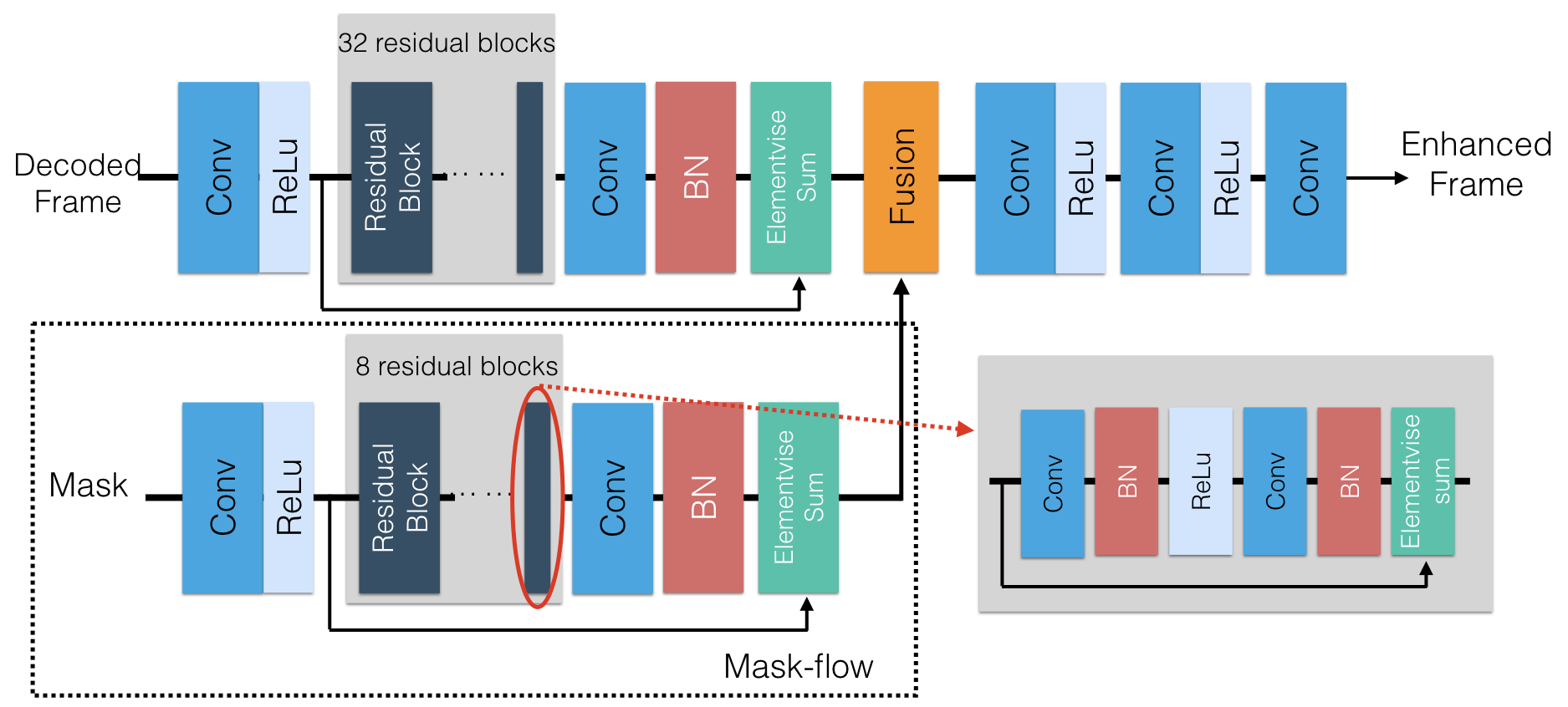}\label{fig:2a}}
  \hfill
  \subfloat[]{\includegraphics[width=0.28\textwidth,height=0.23\textwidth]{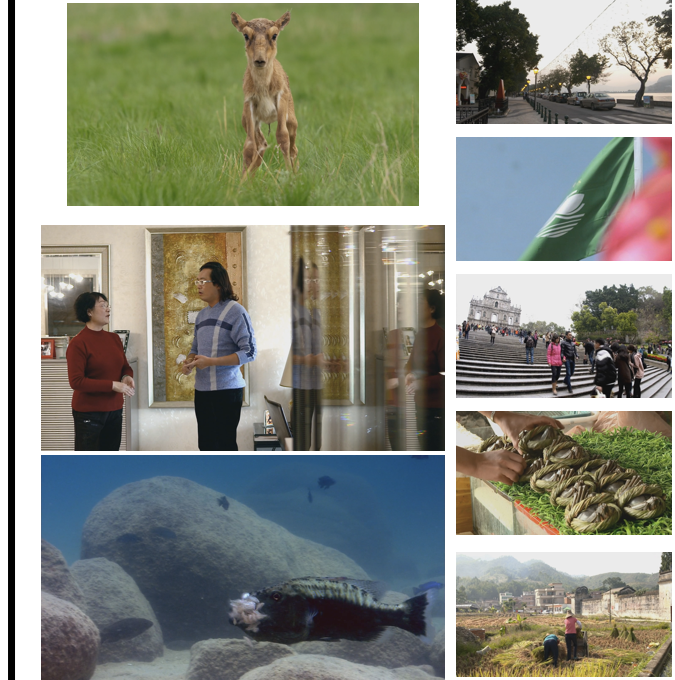}\label{fig:2b}}
  \hfill
  \caption{(a) Double-input convolutional neural network with add-based fusion strategy. (b) Example snapshots of our dataset.}
  \label{fig:base_network}
\end{figure*}

\subsection{Double-input convolutional neural network}
\label{sec:base_network}

The proposed double-input convolutional neural network integrates partition information with add-based fusion strategy and enhances the quality of compressed frames. Its architecture is shown in Fig. \ref{fig:2a}. This CNN contains two streams in the feature extracting stage so as to extract features for the decoded frame and its corresponding mask, respectively. Each residual block \cite{resnet,block,srgan} in the feature extracting stage has two convolutional layers with 3$\times$3 kernels and 64 feature maps, followed by batch-normalization \cite{bn} layers and ReLu activation functions. Then, the feature maps of the mask and decoded frame are fused by the add-based fusion strategy and are fed to the rest three convolutional layers. These three layers with 3$\times$3 kernels and 64 feature maps are utilized for feature enhancement, mapping, and reconstruction as described by \cite{arcnn}. When training the network, the Mean Squared Error between the original raw frame and the CNN output is used as the loss function.

Compared with the existing compressed video enhancement methods \cite{vrcnn,qecnn}, our network has two differences: (1) We introduce two stream inputs to include both the decoded frame and the partition information. (2) We use a residual architecture to perform the feature extraction. The deep residual stream can capture the feature of input in a more distinctive and stable way.

\section{Experimental Results}
\label{sec:ex}
\subsection{Dataset \& experimental settings}
\textbf{Dataset}. In order to construct a reliable double-input CNN, we establish a large-scale dataset. The dataset is derived from 600 video clips with various resolutions. Fig. \ref{fig:2b} shows some snapshots of the video clips. All raw video clips are encoded by HM-16.0 at Low-delay P \cite{schwarz2005hierarchical} at QP=22, 27, 32, and 37. In each raw clip and its compressed clip, we randomly select 3 raw frames and the corresponding decoded frames to form 3 training frame pairs. For each frame pair, we divide them into 64$\times$64 sub-images without overlap resulting 202,251 sub-image pairs.

\begin{table}[t]
\centering
\caption{Comparison of different mask and fusion methods on $\Delta$PSNR (dB) over HM-16.0 baseline at QP=37}
\footnotesize{
\label{table1}
\setlength\tabcolsep{1.5pt}
\begin{tabular}{|c|c|M{9.7mm}|M{9.7mm}|M{9.7mm}|M{9.7mm}|M{9.7mm}|}
\hline
  Class & Sequence & 1-in & 2-in\newline+BM \newline+AF &2-in\newline+MM\newline+EF&2-in\newline+MM\newline+CF & 2-in\newline+MM\newline+AF\\
\hline
 \multirow{4}{*}{A} & Traffic & 0.31 &0.37&0.36& 0.33& \textbf{0.39}\\
 \cline{2-7}
 & PeopleOnStreet& 0.56& \textbf{0.64}& \textbf{0.64}&0.61&\textbf{0.64}\\
 \cline{2-7}
  & Nebuta& 0.27& 0.20 & 0.25 &0.30&\textbf{0.32}\\
 \cline{2-7}
  & SteamLocomotive& 0.19& 0.12 & 0.18& 0.19&\textbf{0.22}\\
 \cline{1-7}
  \multirow{5}{*}{B}& Kimono& 0.36& 0.39 & 0.38& 0.39&\textbf{0.41}\\
 \cline{2-7}
  & ParkScene& 0.17& 0.19 & 0.19& 0.19&\textbf{0.20}\\
 \cline{2-7}
  & Cactus& 0.23 & 0.31 & 0.31& 0.27&\textbf{0.34}\\
 \cline{2-7}
  & BQTerrace& 0.18& 0.29 & 0.28& 0.29&\textbf{0.38}\\
 \cline{2-7}
  & BasketballDrive& 0.19& 0.32 & 0.31 & 0.30&\textbf{0.35}\\
 \cline{1-7}
  \multirow{4}{*}{C}& RaceHorses& 0.26& \textbf{0.30} & 0.29 & 0.29&0.29\\
 \cline{2-7}
  & BQMall& 0.10& 0.25 & 0.23 & 0.27&\textbf{0.36}\\
 \cline{2-7}
  & PartyScene& 0.11& 0.17 & 0.15 & 0.19&\textbf{0.27}\\
 \cline{2-7}
  & BasketballDrill& 0.22& 0.34 & 0.32 & 0.32&\textbf{0.47}\\
 \cline{1-7}
  \multirow{4}{*}{D}& RaceHorses& 0.31& \textbf{0.42}& 0.41& 0.41&0.41\\
 \cline{2-7}
  & BQSquare& -0.04& 0.22& 0.16& 0.24&\textbf{0.50}\\
 \cline{2-7}
  & BlowingBubbles& 0.13& 0.22& 0.20& 0.22&\textbf{0.26}\\
 \cline{2-7}
  & BasketballPass& 0.19& 0.35& 0.32& 0.36&\textbf{0.40}\\
 \cline{1-7}
  \multirow{3}{*}{E}& FourPeople& 0.44& 0.55& 0.54& 0.53&\textbf{0.62}\\
 \cline{2-7}
  & Johnny& 0.35& 0.48& 0.47& 0.45&\textbf{0.54}\\
 \cline{2-7}
  & KristenAndSara& 0.39& 0.56&0.52& 0.52&\textbf{0.59}\\
 \cline{1-7}
  \multicolumn{2}{|c|}{\textbf{Average}} & 0.25& 0.33 & 0.32& 0.33&\textbf{0.40}\\
 \cline{1-7}
\hline
\end{tabular}}
\end{table}

\textbf{Experimental settings}. We implement the proposed model using TensorFlow \cite{tensorflow}.
During training, we use a mini-batch size of 32. We start with a learning rate of 1e-04, decay the learning rate with a power of 10 at the 20th epochs, and terminate training at 40 epochs. An individual CNN is trained for each QP. In order to save training time, we first train the double-input CNN at QP=37 from scratch and the other networks at QP=32, 27, 22 are fine-tuned from it.


During the evaluation, we test our trained model on 20 benchmark sequences from the common test conditions of HEVC \cite{ctc}. The performance of quality enhancement is measured by PSNR improvement ($\Delta$PSNR) and the Rate-distortion performance is measured by the Bjontegaard Distortion-rates (BD-rate) \cite{bd-rate} savings over HM-16.0 baseline. Similar to existing works, the performances on Y-channel are evaluated in our experiments.

\subsection{Results on different mask generation \& mask-frame fusion strategies.}
Table \ref{table1} compares the performance of different mask generation and mask-frame fusion strategies described in Section \ref{sec:mask_and_fusion}. In Table \ref{table1}, \emph{1-in} represents a single-input baseline of our approach where the mask-flow input is deleted from the framework of Fig. \ref{fig:2a}; \emph{2-in+MM+AF} represents our double-input CNN using the mean-based mask and add-based fusion strategy. Note that the performances of all methods are evaluated by the PSNR gain over HM-16.0 baseline at QP=37. From Table \ref{table1}, we can have the following observations:
\begin{enumerate}[noitemsep]
\item When looking at different mask generation strategies, the boundary-based mask strategy (2-in+BM+AF) cannot provide noticeable improvement (0.08 dB over \emph{1-in}). This is because only marking boundary pixels in a mask is less effective in highlighting the partition modes in a frame. Comparatively, the mean-based mask (2-in+MM+AF) can obtain more obvious PSNR improvement (0.15 dB over \emph{1-in}). This indicates its effectiveness in capturing the partition modes in a frame.

\item As for mask-frame fusion strategies, the add-fusion strategy (2-in+MM+AF) can obtain a large PSNR gain of 0.4 dB. This shows the effectiveness of the proposed fusion strategy. Comparatively, the concatenate-fusion (2-in+MM+CF)  and early-fusion (2-in+MM+EF) strategies obtains fewer gains. This is probably because these fusion strategies are less compatible with the CNN model used in this paper. Their performances may be more obvious when combined with other CNN models.

\item The best performance is obtained when using mean-based mask and add-fusion (2-in+MM+AF), which can obtain over 0.15 dB improvement over single-input method. This indicates that when strategies are properly selected, introducing partition information is indeed useful to improve the quality of compressed videos.
\end{enumerate}

\subsection{Comparison with the existing methods}
Table \ref{table2} further compares the overall BD-rate saving \cite{bd-rate} of different methods over the standard HEVC test model (HM-16.0).
Five methods are compared in Table \ref{table2}: (1) VRCNN \cite{vrcnn} which is a benchmark CNN-based compressed-video enhancement method; (2) QECNN-P \cite{qecnn} which is a state-of-the-art compressed-video enhancement method for P frames in HEVC; (3) Our (1-in), which is the single input baseline of our approach; (4) VRCNN+MM+AF, which integrates our partition-mask-based approach into the existing VRCNN method; (5) Our (2-in+MM+AF), which is the full version of our approach with mean-based mask and add-based fusion. Note that in order to have a fair comparison, all methods are trained using the same dataset (i.e., our dataset) and evaluated under the same setting. From Table \ref{table2}, we can observe that:
\begin{enumerate}[noitemsep]
\item The full version of our approach (\emph{our+2-in+MM+AF}) achieves the best performance overall the compared methods. Specifically, it can obtain over 9.76\% BD-rate reduction from standard HEVC and 4\% BD-rate reduction when compared with the state-of-the-art QECNN method. This clearly indicates the effectiveness of our partition-mask-based approach.

\item When integrating our partition-mask strategy,  the \emph{VRCNN+MM+AF} can also obtain 3\% BD-rate improvement over the original VRCNN method. This demonstrates that our partition-mask-based approach can be easily combined with the existing methods to provide further improved methods.

\item Our baseline single-input method (\emph{our+1-in}) can also obtain satisfactory results when compared with the existing methods (VRCNN, QECNN-P). This implies that the baseline CNN model used in our approach is effective in handling the visual information of the input decoded frames.




\end{enumerate}

\begin{table}[t]
\centering
\caption{Comparison of different methods on BD-rate (Y,\%) saving over HM-16.0 baseline}
\footnotesize{
\label{table2}
\setlength\tabcolsep{1.5pt}
\begin{tabular}{|c|c|c|c|M{9mm}||M{10mm}|M{9mm}|}
\hline
 Class & Sequence & VRCNN & QECNN-P & Our\newline(1-in) & VRCNN\newline+MM\newline+AF & Our \newline(2-in\newline+MM\newline+AF) \\
\hline
  \multirow{4}{*}{A} & Traffic & -6.84 &-8.28& -9.271&-9.09&\textbf{-11.35}\\
 \cline{2-7}
  & PeopleOnStreet& -7.41& -8.66& -9.84 & -9.43&\textbf{-10.36} \\
 \cline{2-7}
  & Nebuta& -5.65& -7.56& -6.23 &-6.55 & \textbf{-7.85} \\
 \cline{2-7}
  & SteamLocomotive& -7.71&-9.18 & -10.22 & -9.89&\textbf{-10.6}\\
 \cline{1-7}
  \multirow{5}{*}{B}& Kimono& -7.39&  -8.70& -9.49 & -9.07&\textbf{-10.91} \\
 \cline{2-7}
  & ParkScene& -3.97& -4.73& -5.4 & -5.32& \textbf{-6.92}\\
 \cline{2-7}
  & Cactus& -5.86& -7.39& -8.13 &-8.16 &\textbf{-10.53} \\
 \cline{2-7}
  & BQTerrace& -1.73& -4.87& -7.25 & -6.99&\textbf{-11.07}\\
 \cline{2-7}
  & BasketballDrive& -3.75&-5.91 & -6.42 & -6.74 &\textbf{-11.10}\\
 \cline{1-7}
  \multirow{4}{*}{C}& RaceHorses& -3.6&-4.78 & -5.57 & -5.44& \textbf{-6.45}\\
 \cline{2-7}
  & BQMall& 0.11& -2.91& -4.01 & -3.97&\textbf{-7.62}\\
 \cline{2-7}
  & PartyScene& 2.72& -1.03& -2.48 & -2.08&\textbf{-4.84}\\
 \cline{2-7}
  & BasketballDrill& -0.08& -2.36& -5.71 & -4.64& \textbf{-10.65}\\
 \cline{1-7}
  \multirow{4}{*}{D}& RaceHorses& -4.05& -5.03& -6.66 & -6.41&\textbf{-7.58}\\
 \cline{2-7}
  & BQSquare& -0.57& -0.11& -2.48 &-2.72 &\textbf{-8.48}\\
 \cline{2-7}
  & BlowingBubbles& -0.15& -2.07& -4.12 &-3.43 & \textbf{-6.33}\\
 \cline{2-7}
  & BasketballPass& -0.15& -2.37& -4.49 & -4.02& \textbf{-7.73}\\
 \cline{1-7}
  \multirow{3}{*}{E}& FourPeople& -7.12& -9.27& -10.69 & -10.33& \textbf{-13.91}\\
 \cline{2-7}
  & Johnny& -7.00& -9.78& -10.40 &-11.41 &\textbf{-17.22} \\
 \cline{2-7}
  & KristenAndSara& -7.13& -9.21& -9.5 & -10.56&\textbf{-13.78} \\
 \cline{1-7}
 \multicolumn{2}{|c|}{\textbf{Average}} & -3.81& -5.71& -6.92 & -6.81 & \textbf{-9.76}\\
\hline
\end{tabular}}
\end{table}

\section{Conclusion}
This paper presents a novel approach for enhancing compressed videos in HEVC. Our approach utilizes the partition information already existing in the bitstreams to design a mask and integrate it with the decoded image in CNN to guide the frame quality enhancement process. Experimental results show that our approach is more effective in handling the visual quality degradation introduced by HEVC encoder, and thus obtaining the best post-processing performance. Furthermore, it can also be applied to the existing compressed-video enhancement methods and bring further improvement.
\label{sec:conclu}

\section*{Acknowledgments}
This work is supported in part by NSFC (61471235), Shanghai `The Belt and Road' Young Scholar Exchange Grant (17510740100).

\bibliographystyle{IEEEbib}
\bibliography{refs}

\end{document}